\documentclass[a4paper,11pt]{article}
\pdfoutput=1 % if your are submitting a pdflatex (i.e. if you have
\usepackage{jcappub} % for details on the use of the package, please
\usepackage[T1]{fontenc} % if needed

\usepackage{txfonts}
\usepackage{pstricks}
\usepackage{graphicx}
\usepackage{graphbox}
\usepackage{subfig}
\usepackage{bm}% bold math

\usepackage{booktabs}
\usepackage{diagbox}
\usepackage{multirow}

\let\vec\mathbf

\title{Experimental hint for gravitational CP violation}

\author{V.~Gharibyan}

\affiliation{Deutsches Elektronen-Synchrotron DESY, D-22603 Hamburg, Germany}

\abstract{
An equality of particle and antiparticle gravitational interactions
holds in general relativity and is supported by indirect 
observations. Gravity dependence on rotation or spin direction
is experimentally constrained for non-relativistic matter.
Here a method based on high energy Compton scattering is developed
to measure the gravitational interaction of accelerated charged particles.
Within that formalism the Compton spectra measured at HERA 
rule out the speculated anti-gravity possibility for antimatter
at a confidence level close to $100\%$. 
The same data, however, imply a gravitational CP violation around 
13~GeV energies, by a maximal amount of $(9 \pm 2) \cdot 10^{-12}$ for 
the charge and $(13\pm 3)\cdot 10^{-12}$ for the space parity.
The detected asymmetry hints for a stronger gravitational coupling to left 
helicity electrons relative to right helicity positrons.}

\begin{document}
\maketitle
\flushbottom


\section{Introduction}
Extreme weakness of gravitation makes it the 
least experimentally investigated fundamental interaction
at sub-atomic scales, with elementary particles.
Any test with a single carrier of the gravitational 
force, hypothetical graviton, is unfeasible whereas 
in the electroweak or strong interactions  experimental
results with a photon, W, Z boson or a gluon are readily available
at sufficiently high energies.
The weak gravity combined with a rarity and 
vulnerability of antiparticles drives any attempt of testing
the antimatter gravitation to its technical limits.
Einstein's general  
relativity~\cite{Einstein-GR}, the currently accepted 
theory of gravitation, does not distinguish
between particles and antiparticles or their properties
such as spin or helicity. The general relativity is based on
universality of gravity and equivalence principle, and,  
there is no alternative theory available to predict or 
describe violations of these principles in a consistent way. 
Departures from perfect spin or particle-antiparticle 
symmetry are allowed in some quantum gravitation 
scenarios~\cite{AmelinoCamelia:2008qg}. 
Hence, observations of antiparticle gravitation could serve 
as an experimental input for quantum gravity~\cite{Nieto:1991xq}.
Additional motivations for such investigation are the
still unexplained matter-dominant universe~\cite{Agashe:2014kda} and the
connection of antimatter's possible anti-gravity~\cite{Villata:2011bx} 
to the accelerated expansion of the universe~\cite{Riess:1998cb}. 
One can also think about a possible particle-antiparticle gravitational 
asymmetry and helicity dependence from an analogy to electroweak interactions, 
where a photon's massive partners, W and Z bosons, are considered responsible
for space and charge parity violations~\cite{Beringer:1900zz}. 
Thus, possible massive or lower spin gravitons could introduce similar 
violations~\cite{Goldhaber:2008xy} that may remain hidden at low energies 
and will become detectable at high energies.

Another possibility to incorporate asymmetric space or charge 
gravitational interactions is a modification of vacuum properties described
within the Standard Model Extension (SME)~\cite{Kostelecky:2010ze}
by Lorentz violating terms in Lagrangian. Such action based approach is further 
developed for anomalous antimatter gravity  in~\cite{Kostelecky:2015nma}.

Indirect observations of matter-antimatter gravitational asymmetry 
involve nuclei with different content of quark-antiquarks  
in the equivalence principle E\"otv\"os type 
experiments~\cite{Schlamminger:2007ht,Adelberger:2009zz}.  
Using $CPT$ conservation the observed stringent limits for the 
equivalence principle violating matter could be expanded to 
a limit below $10^{-7}$ for the matter-antimatter low energy 
gravitational asymmetry~\cite{Alves:2009jx}.  
Hughes and Holzscheiter have set limits on the matter-antimatter 
gravitational differencies from an equality of particle-antiparticle 
cyclotron frequencies~\cite{Hughes:1990ay}. As the source of gravitation, 
however, they have used the potential of the local Supercluster while for 
the Earth's potential the derived restrictions are completely lifted.
Technical difficulties for  charged antiparticle's 
gravitational coupling's direct measurements
turned physicists' attention to  neutral antimatter 
tests~\cite{Scampoli:2014tpa,  Amole:2013gma, Gabrielse:2012xe} 
which may  deliver conclusive results soon.
The ongoing experiments, however, are still at low energy, and massive
gravitons' interactions may remain unseen.

Spin dependent gravitation (for a review see~\cite{Ni:2009fg}) motivated by 
Lorentz violation~\cite{Kostelecky:2004pd}, 
torsion gravity~\cite{Hehl:1976kj},
exchange of pseudoscalar bosons~\cite{Moody:84}
or few other hypothesis has been constrained by low 
energy spin-polarized experiments, such as
the test with polarized torsion pendulum~\cite{Heckel:2008hw}.
Here also the high energy could reveal gravitation's preference for a helicity  
which is hidden at low energies.

In this Letter I will demonstrate an extreme sensitivity of a high energy 
process - laser Compton scattering - to an antiparticle's  hypothetical
anti-gravity and gravitational charge and space parity violation.
Next, applying the developed formalism to the 
existing data of the HERA Compton polarimeter, I will compare
the $\gamma$-spectra  generated by 
electrons and positrons to measure the charge and spin asymmetry 
for their gravitational interaction.
Systematic effects and prospects for other tests will be discussed at 
the end.

\section{Refraction in gravitational field}

In an earlier publication,  high energy Compton scattering 
sensitivity has been shown to a Planck scale dispersive and birefringent vacuum 
model~\cite{Gharibyan:2012gp}. 
Subsequently, I applied the same formalism to the Earth's gravity
assuming the real gravitational field's induced refractivity only for the
Compton photons~\cite{Gharibyan:2014mka}. 
The refraction, however,  also affects the 
leptons  involved in the scattering~\cite{Evans:2001hy} in agreement 
with the equivalence principle.
This makes the ref.~\cite{Gharibyan:2014mka} conclusions about 
the general relativity violation invalid~\cite{khaladgyan}.

Here I follow the formalism developed by Evans et al.~\cite{Evans:2001hy} 
to find a massive particle's energy-momentum or refraction relation in
a static and isotropic gravitational field described by the Schwarzschild metric.
Combining the Eq.(3) and Eq.(30) from the reference \cite{Evans:2001hy}, 
for a particle at a distance R from mass M, in a weak gravitational field
$GM/R \ll c^2$ one can derive a refraction relation 
\begin{equation}
c\frac{P}{\cal E} = \frac{v}{c} + \frac{2 G M}{c^2 R},
\label{disp}
\end{equation}
where $G$ is the gravitational constant, $c$ is the speed of light and 
\hbox{${\cal E},P,v$} are  energy, momentum,  velocity of the particle.
This relation is also valid  for massless particles: at
$v=c$ it describes the photon refraction in a gravitational field
in a form derived by many authors; see ref.~\cite{de-Felice:1971ui} 
and references therein, or, for a more recent reference, see ref.~\cite{Sen:2010zzf}.
%%<<<<<<<<<<<<<<<<<<<<<<<<<<<<<<<<<<<<<<<<<<<<<<<
The second term in Eq.(\ref{disp}) is a scaled gravitational
potential \hbox{$U(G,M,R)=-GM/R$} of the gravitating mass.
This term can be dropped from Eq.(\ref{disp}) unless there is 
a change in the potential
\begin{equation}
\Delta U = U\frac{\Delta G}{G}+ U\frac{\Delta M}{M} - U\frac{\Delta R}{R}.
\label{du}
\end{equation}
Indeed, at $\Delta U=0$ the physical processes (involving one or more particles) 
are not altered by the extra gravitational term in Eq.(\ref{disp}) since the 
potential $U$ cancels in energy-momentum conservation. 
Situation changes with $\Delta U \neq 0$. Thus, for $\Delta R \neq 0$ case, 
differentiated Eq.(\ref{disp}) describes particle's gravitational energy change 
(frequency red/blue shift for $v=c$) or deflection - using the 
principle of least action.
 While these are classical gravity processes validated by
multiple  measurements, the first term in Eq.(\ref{du}), 
\hbox{$\Delta U_G = U\Delta G/G$}, violates 
equivalence principle and has intensively been constrained by pendulum null 
experiments with different materials - all for the non-relativistic ($c=\infty$) case 
(see ref.\cite{Schlamminger:2007ht} and references therin).
Thus, Eq.(\ref{disp}) follows from general relativity in a weak field 
(metric is replaced by gravitational potential) and     
describes relativistic gravitational effects in accord with the 
equivalence principle.

To allow departure from the equivalence principle let us 
retain the interaction strength $G$ for matter particles 
and use a different strength $G_p$ for antimatter leptons  
to write  Eq.(\ref{disp}) for positrons in the 
following form
\begin{equation}
c\frac{P}{\cal E} =\frac{v}{c} - 2\frac{U}{c^2}\biggl(1+\frac{\Delta G_p}{G}\biggr),
\label{disp1}
\end{equation}
with \hbox{$\Delta G_p = G_p - G$}.
For an anti-gravitating positron \hbox{$G_p=-G$}.

In a similar manner, within the spin affected gravitation, we can assign an interaction 
constant $G_{-}$ to particles possessing a negative(left) helicity and modify the 
Eq.(\ref{disp}) to include a helicity dependent term \hbox{$\Delta G_{-} = G_{-} - G$}
\begin{equation}
c\frac{P}{\cal E} =\frac{v}{c} -2\frac{U}{c^2}\biggl(1+\frac{\Delta G_{-}}{G}\biggr).
\label{disp2}
\end{equation}
 Positive(right) helicity particles' coupling constant $G_{+}$ is assumed to
provide an average $(G_{-}+G_{+})/2=G$ for unpolarized or zero
helicity case, so that $\Delta G_{-} = -\Delta G_{+} $.

Anomalous dispersion or refraction relations similar to the Eq.(\ref{disp1}) 
or Eq.(\ref{disp2}) are often 
applied for vacuum, within the context of action based 
Lorentz violation~\cite{Tasson:2014dfa,Liberati:2013xla}. 
Such models have an advantage to access modified dynamic properties (cross-sections) 
of considered processes though stringent observational and experimental limits, 
in particular from a Compton scattering test~\cite{Bocquet:2010ke}, exist
for Lorentz violating vacuum~\cite{Kostelecky:2008ts}.

Here we limit ourselves exceptionally with kinematics in a real gravitational field to
investigate how the hypothetical refractions by  Eq.(\ref{disp1}) and Eq.(\ref{disp2})
are affecting the high energy laser Compton process.
%%%%==================================================================

Similar to Eq.(\ref{disp1}) assumption has recently been considered in hypothetical 
vacuum Cherenkov radiation formalism to constrain charge parity violation in 
gravitational field~\cite{Kalaydzhyan:2015ija}.

\section{The Compton process affected by gravity}

Using energy-momentum conservation with  Eq.(\ref{disp}) and  Eq.(\ref{disp1}), 
when in the Earth's gravitational field a photon scatters off a 
positron with energy ${\cal E}$, 
the Compton scattering kinematics is given by (from here on natural units are assumed)
\begin{equation}
{\cal E}x - \omega (1+x+\gamma^2 \theta^2) + 
4\omega \biggl(1 - \frac{\omega}{{\cal E}} \biggr) \gamma^2 U
\frac{ \Delta G_p }{ G} = 0,
\label{comp0}
\end{equation}
where \hbox{$x=4\gamma \omega_0\sin^2{(\theta_0/2)}/m$}, 
with $m$ and $\gamma={\cal E}/m$  being the mass and Lorentz factor of the
initial positron, respectively. The initial photon's energy and angle are denoted
by $\omega_0$ and $\theta_0$, while the refraction of  Eq.(\ref{disp}) is in effect 
for the scattered photon with energy $\omega$ and angle $\theta$; the angles are 
defined relative to the initial positron.
This kinematic
expression is derived for weak gravity and high energies,  
i.e., the  $\mathcal{O}(U^2)$, $\mathcal{O}(\theta^3)$, and 
$\mathcal{O}(\gamma^{-3})$ terms are neglected.
In this approximation the refraction effect of Eq.(\ref{disp}) for the initial laser 
photon is negligible. For the initial and scattered positrons the refraction
of Eq.(\ref{disp1}) is applied.
To determine the outgoing photon's maximal energy,  Eq.(\ref{comp0}) is solved 
for $\omega$ at $\theta=0$ with the following result:
\begin{equation}
\omega_{max} = {\cal E}~\frac {b+q-\sqrt{b^2+q(q-2b+4)} } {2\,q},
\label{comp}
\end{equation}
where $b=1+x$ and $q= -2\gamma^2 U \Delta G_p / G$.
Thus, in high energy Compton scattering
the small factor $ U \Delta G_p / G$  is amplified by $\gamma^2$, 
allowing one to measure it by detecting the extreme energy of the scattered photons 
$\omega_{max}$, or positrons \hbox{${\cal E}-\omega_{max}$} (Compton edge).  

\begin{figure}[h]
\centering
\includegraphics[scale=0.47]{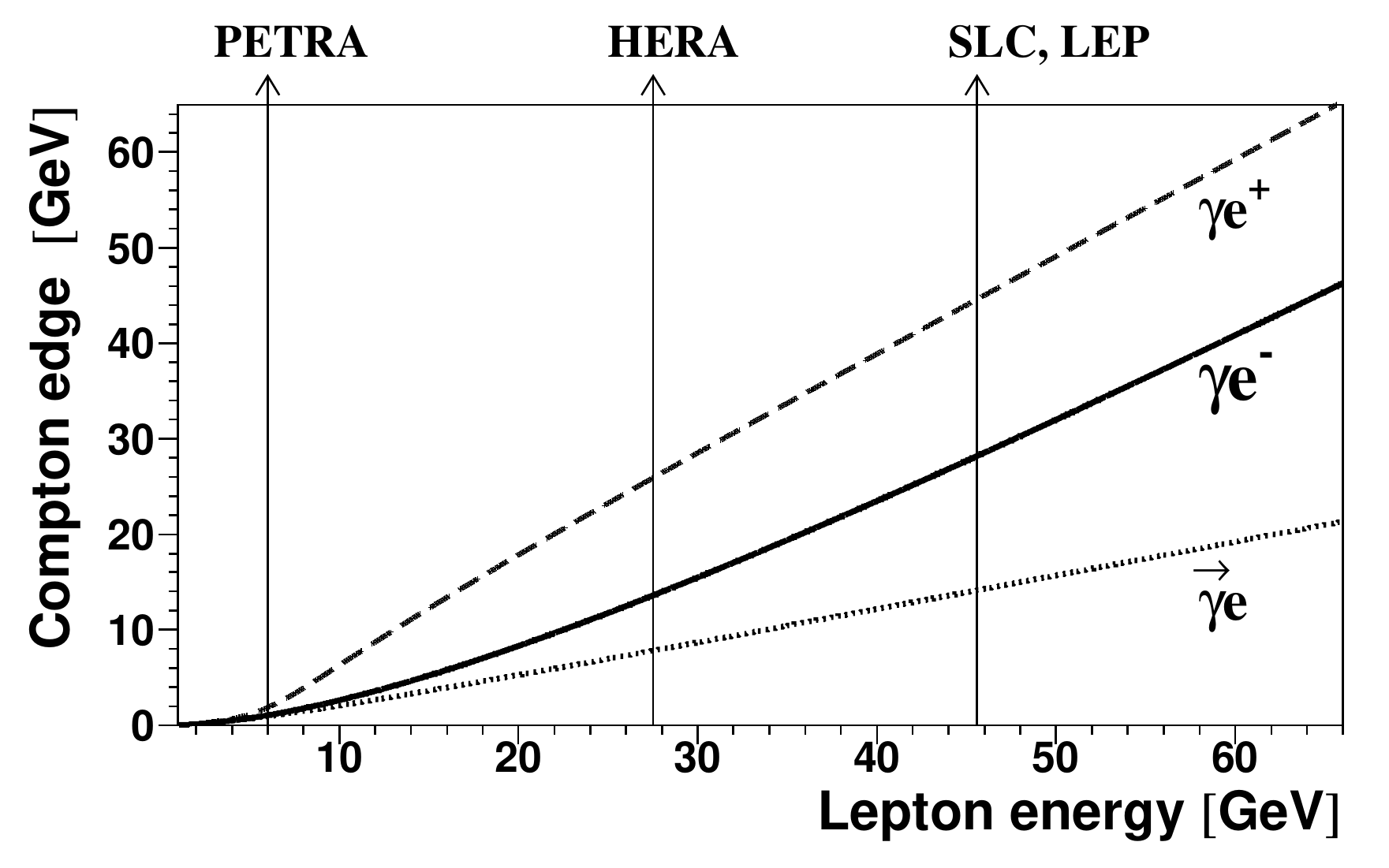}
\caption{\label{fig1}
The maximal energy of Compton scattered photons (Compton edge)
and its dependence on the initial lepton energy 
for a head-on collision with 532nm laser light, in the Earth's 
gravitational field.
The solid line is for matter gravity ($e^-$, electron, $G$) 
while the dashed line shows antimatter anti-gravity case ($e^+$, positron, $-G$).
The dotted line corresponds to gravitationally sterile right (positive helicity) 
Compton photons ($\vec{\gamma}$, $G_+ =0$).
Names of $e^+ e^-$ accelerators are printed at the upper part.
}
\end{figure}

In order to estimate the method's sensitivity, I calculate the Compton edge
from the Eq.(\ref{comp})
for an incident photon energy 2.32~eV (the widely popular green laser) 
in the Earth's gravitational field 
\hbox{($U=-G M_\oplus /R_\oplus = -6.95\times 10^{-10}$)}, 
at different energies of the accelerator leptons.  
The resulting dependencies for a matter (electron) gravity and antimatter 
(positron) anti-gravity are presented in Fig.~\ref{fig1}.
%%%%%%%%%%%%%%%%%%%%%%%%%%%%%%%%%%%%%%%%%%%  fig 11111111111111111111111
%%%%%%%%%%%%%%%%%%%%%%%%%%%%%%%%%%%%%%%%%%%%%%%%%%%%%%%%%%%%%%%%%%%%%%%%%%%%%%%%%%%
The plot shows considerable sensitivity, which grows toward high energies
in a range available to  accelerating laboratories.
For handling  measurement's systematic errors, from an experimental point of view,
it is more precise to measure a relative asymmetry rather than absolute Compton 
edge energy. 
Therefore, we form an asymmetry of Compton edges measured on positrons
($\omega_{max}^p$) and electrons ($\omega_{max}^e$)
\begin{equation}
A=\frac{\omega_{max}^p-\omega_{max}^e}{\omega_{max}^p+\omega_{max}^e}
\label{asym}
\end{equation}
and use Eq.(\ref{comp0}) to find the charge parity gravitational violation 
magnitude
\begin{equation}
\Delta U_{Gp} \equiv U\frac{\Delta G_p}{G}=-\frac{1}{2 \gamma^2}\cdot
\frac{A(1-A)(1+x)^2}{(1+A)(2Ax+A-1)}.
\label{dgog}
\end{equation}

For evaluating a sensitivity of the laser Compton process to the 
spin dependent  gravitation let's investigate left helicity photon  
scattering off unpolarized (zero mean helicity) leptons' beam. According 
to helicity conservation, the scattered gamma-quantum will 
retain the helicity of the initial photon at Compton 
edge energies~\cite{mcmaster:1961xe}. 
Hence we apply the helicity modified refraction relation of  
Eq.(\ref{disp2}) to the initial and final photons, reserving 
the refraction of Eq.(\ref{disp}) for the initial lepton.     
After scattering the leptons become longitudinally polarized acquiring 
\hbox{$P_l=x(2+x)/(1+(1+x)^2)$} portion of the initial photon 
circular polarization~\cite{Kotkin:2002ra, Lipps}. 
Therefor, for the scattered lepton, the term $\Delta G_{-}/G$ in 
energy-momentum relation (\ref{disp2}) must be scaled by a factor $-P_l/2$,
where the $-1/2$ coefficient counts for half spin and positive helicity of
the lepton.       
This modifications will bring the Eq.(\ref{comp}), Eq.(\ref{dgog}) for the 
charge parity case to 
\begin{equation}
\omega_{max}^- = {\cal E}~\frac {\sqrt{b^2+4u(u+x+P_l-1)}-b+2u(1+P_l)}{2u(2+P_l)},
\label{comp_}
\end{equation}
and, with a measured energy asymmetry
\hbox{$A_- =(\omega_{max}^--\omega_{max}^0)/(\omega_{max}^-+\omega_{max}^0)$},  
to 
\begin{equation}
\Delta U_{G-} =\frac{1}{ 2\gamma^2}\cdot 
\frac{2A_-(1-A_-)(x^2+2x+2)(1+x)^2}{A_-^2x_1+A_-x_2-(2x^2+3x+2)},
\label{dgog_}
\end{equation}
for the space parity case. 
In the above equations the upper index ($-,0$) stands for the scattered 
photon's average helicity and the following assignments are made:
\hbox{$u= -2\gamma^2 U \Delta G_- / G $},
\hbox{$\Delta U_{G-}=U \Delta G_- / G$},
\hbox{$x_1=8x^3+22x^2+21x+6$},
\hbox{$x_2=4(x^3+x^2-0.5x-1)$}. 
Energy dependence of  Eq.(\ref{comp_}),
if the Earth's gravity attracts only left helicity particles,   
is plotted on the Fig.~\ref{fig1}. From the plot and formulas we can conclude
that Compton process sensitivity to the gravitational space parity violation
is sufficient for introducing a Compton edge sizable shift in a scenario of
the helicity dependent gravitational field. 

For estimating the sensitivity in Fig.~\ref{fig1}
the Earth's gravitational potential $6.95\cdot 10^{-10}$ is used. 
There is though a much more greater gravitational field 
at the Earth's surface, the largest, local Supercluster's 
potential \cite{Hughes:1990ay} with a magnitude $3\cdot 10^{-5}$.  
For gravitational effects induced by the third term proportional to 
$\Delta R/R$ in Eq.(\ref{du}), the Earth's potential prevails
(due to large $1/R$ factor) while in our case, for the
first term in Eq.(\ref{du}) with $\Delta G/G$, the local Supercluster
potential contribution is dominant.  Exact potential selection, however,
is important only when separating the  $\Delta G/G$ contribution from 
the $\Delta U$ factor. 
The Compton scattering could measure only the factors
 $\Delta U_{Gp}$ or $\Delta U_{G-}$.
  
Besides the gravitation, other interactions also may alter 
energy-momentum relation~\cite{Latorre:1995cv,Dittrich:1998fy}.
For the Compton process a background electromagnetic field 
is the main competitor to gravity, 
and, since electric fields are usually shielded by conductive 
beam pipes, let's estimate influence of magnetic fields.
According to Eq.(1.1) from the reference~\cite{Latorre:1995cv}, 
the maximal impact to energy-momentum relation for photons 
in a magnetic field $B$ is presented by
\begin{equation}
\frac{P}{\cal E} = 1 - \frac{14}{45} \frac{\alpha^2}{m^4} B^2 ,
\label{emref}
\end{equation} 
\noindent 
where $\alpha$ is the fine-structure constant.
Hence, refractivity created by a 4~T superconducting magnet
is about $8\cdot 10^{-21}$ and that of Earth's  
magnetic field is $10^{-25}$. These tiny effects are 
still experimentally unreachable, and, compared to 
the gravitational potential magnitude 
(either the Supercluster's $\sim 10^{-5}$ or the Earth's $\sim 10^{-9}$) 
in the Eq.(\ref{disp}), are completely negligible.
  
\section{Experimental results}

\begin{figure}[hb!]
\centering
\includegraphics[scale=0.49]{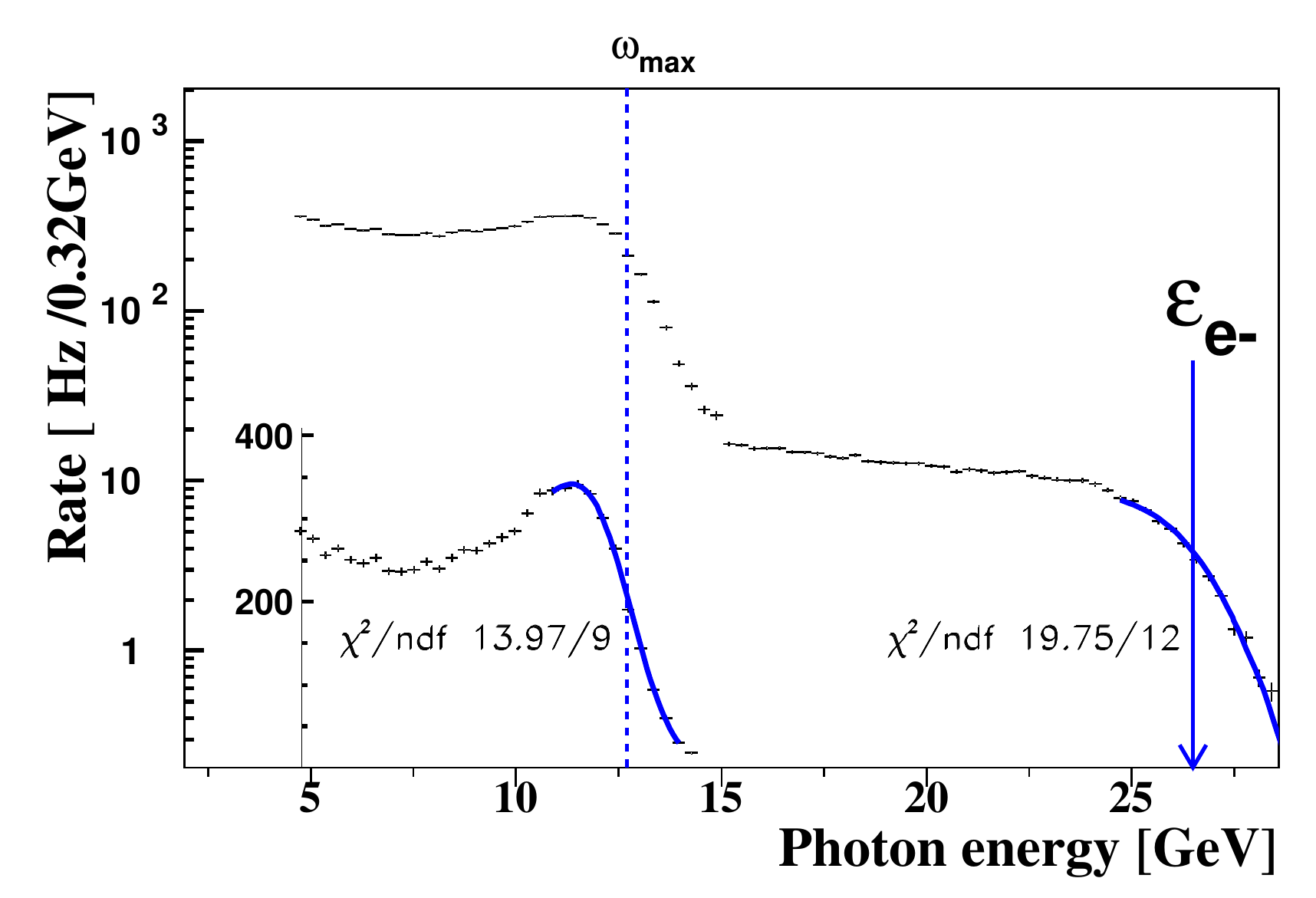}
\caption{\label{fig2}
HERA polarimeter Compton $\gamma$-spectrum 
produced by laser scattering on $26.5~GeV$ electrons,
on top of background Bremsstrahlung with fit results.
The inset displays the background subtracted Compton spectrum.
Vertical  lines show measured values of  the Compton ($\omega_{max}$) 
and Bremsstrahlung (${\cal E}_{e-}$) 
maximal energies. }
\end{figure}

The high-energy accelerators where laser Compton facilities have been operated 
for years, are listed on the upper energy scale of  Fig.~\ref{fig1}. 
As can be seen from the plot, 6~GeV storage rings  have low 
sensitivity while the higher energy colliders (HERA, SLC, LEP) have a great
potential for detecting gravity related energy shifts.
This is true for the HERA and SLC Compton polarimeters but not for the LEP 
polarimeter, which has
generated and registered many photons per machine pulse~\cite{LEP-polarimeter}. 
In this multi-photon regime, any shift of the Compton edge is convoluted with the 
laser-electron luminosity and can-not be disentangled and measured separately.
Unlike the LEP, the SLC polarimeter  operated in multi-electron mode and  
analyzed the energies of interacted leptons using a magnetic 
spectrometer~\cite{ALEPH:2005ab}. However, at SLC only the electron beam was polarized, 
and positron data are missing. Hence, we turn to HERA, which has recorded
Compton measurements for both the electrons and the positrons.
At the HERA transverse polarimeter Compton photons are registered by a calorimeter 
in single particle counting mode. A recorded Compton spectrum 
produced by $514.5$nm laser scattering on $26.5~GeV$ electrons,
 from ref.\cite{Barber:1992fc}, 
is shown in Fig.~\ref{fig2} 
%%%%%%%%%%%%%%%%%%%%%%%%%%%%%%%%%%%%%%%%%%%  fig 2222222222222222222
%%%%%%%%%%%%%%%%%%%%%%%%%%%%%%%%%%%%%%%%%%%
superimposed on a background  Bremsstrahlung distribution.
In contrast to  Compton scattering, in the Bremsstrahlung process the momentum 
transfer is not fixed, and any small refractive  effect is smeared out and becomes 
negligible~\cite{Gharibyan:2003fe}.
Hence, following the analysis in  ref.~\cite{Gharibyan:2003fe}, I calibrate the energy 
scale according to the maximal Bremsstrahlung energy which is found by fitting 
a convolution of parent energy distribution ${d\Sigma}/{d\omega}$ with the 
detector response gaussian function,
\begin{equation}
F(E_\gamma)=N\int^{E_m}_{0} 
\frac{d\Sigma}{d\omega}\frac{1}{\sqrt{\omega}} 
\exp\Biggl({\frac{-(\omega-E_\gamma)^2}
{2\sigma_0^2 \omega}\Biggr) d\omega},
\label{eqfold}
\end{equation}
to the Bremsstrahlung spectrum.
$\sigma_0$  and $E_\gamma$ in the fitting function denote the
calorimeter resolution and detected photon's energy respectively
while the normalizing factor  $N$ and maximal energy $E_m$ are 
free fitting parameters.
The fit is performed using MINUIT~\cite{James:1975dr} and the integral (\ref{eqfold}) 
is calculated numerically applying the Gaussian quadrature algorithm. 
The same fitting function with the Bremsstrahlung parent distribution
replaced by the Compton scattering differential cross-section 
${d\Sigma_C}/{d\omega}$ 
is applied to the background subtracted spectrum to find the Compton
edge  at \hbox{$\omega_{max}^{e0}= 12.70\pm 0.02$~GeV}. 
Here the upper indexes denote  (scattered) lepton type, helicity 
and the 0 stands for a spectrum summed over the laser left and right helicities. 
The fit results together with fit quality estimates are shown in  Fig.~\ref{fig2}.
The absolute value of the Compton edge is calculated from 
3 measurements -- Bremsstrahlung and Compton edges $B_{max}$ and $C_{max}$ 
in ADC units, and electron beam energy $E_{beam}$ in GeV -- by a formula
\hbox{ $\omega_{max} = E_{beam} C_{max}/B_{max}$}. 
More details about the analysis and experimental setup can be found 
in the ref.~\cite{Gharibyan:2003fe}.

\begin{figure}[h]
\centering
\includegraphics[scale=0.49]{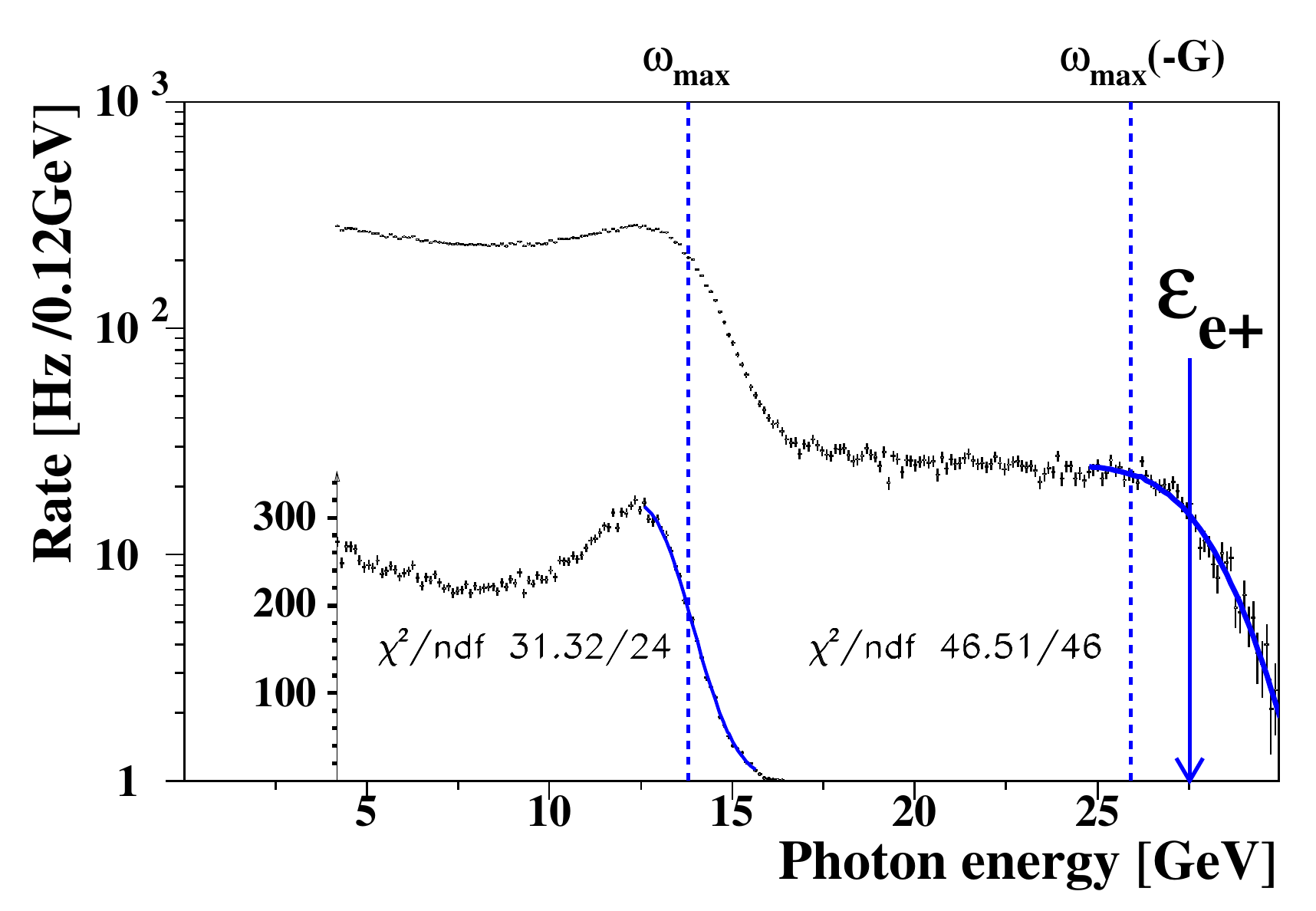}
\caption{\label{fig3}
A similar plot to Fig.~\ref{fig2} for positrons with energy $27.5~GeV$. 
The Compton edge energy for anti-gravitating positrons is indicated by
a vertical line $\omega_{max}(-G)$.}
\end{figure}

The same analysis procedure is applied to a HERA polarimeter 
Compton spectrum that was generated with left helicity photons scattered on
$27.5~GeV$ positrons and 
has been reproduced in Fig.~8 of ref.~\cite{Sobloher:2012rc}. 
As it mentioned above, left helicity laser photons are generating the same (negative) 
helicity scattered positrons at the Compton edge ($\omega_{max}^{p-}$).
The resulting plots with fit quality outcomes are displayed in Fig.~\ref{fig3}.
%%%%%%%%%%%%%%%%%%%%%%%%%%%%%%%%%%%%%%%%%%%  fig 33333333333333333333
%%%%%%%%%%%%%%%%%%%%%%%%%%%%%%%%%%%%%%%%%%%
Comparing the obtained Compton edge  
\hbox{$\omega_{max}^{p-}= 13.80\pm 0.02$}~GeV 
with the photons' maximal energy for the anti-gravitating positrons 
$25.9$~GeV, derived from Eq.(\ref{comp}),  
one can conclude without any advanced systematic error analysis
that anti-gravity for the positrons is ruled out. 
We can estimate accuracy of this statement by assuming an 
absolute energy calibration scale of $0.14~GeV$ as it calculated in
ref.~\cite{Gharibyan:2003fe}. Thus, antimatter's anti-gravity is excluded     
within more than $85\sigma$ confidence interval.

Using only the quoted statistical errors
for the $\omega_{max}^{e0}$, $\omega_{max}^{p-}$ we obtain a
positron-electron and negatve-zero helicity Compton edge asymmetry\\
\hbox{$A_{ep0-}=0.01297 \pm 0.0013.$}
In order to account for the different energies
of accelerated electrons and positrons $26.5$ and $27.5$ GeV, 
the measured maximal energies in the asymmetry
calculation have been normalized to $13.10$ and $13.87$~GeV
for the electrons and positrons, respectively.
These are the expected Compton edge values from  Eq.(\ref{comp}) or Eq.(\ref{comp_})
in the absence of gravitational anomaly, at $\Delta G_p = 0$ or $\Delta G_- = 0$
respectively. 

Since the spectra for electrons and positrons are detected with the
same experimental setup, i.e. with the same laser, geometry and detector,
both measurements will experience the same systematic influences 
that will cancel out or reduce greatly in the asymmetry of Eq.(\ref{asym}). 
Hence, we omit systematic corrections with associated errors 
applied to polarization or energy measurements (as
described in the refs.~\cite{Barber:1992fc,Gharibyan:2003fe,Sobloher:2012rc,Gharibyan:2006note}), 
in order to analyze instrumental errors specific to the detected asymmetry
which we rewrite in a more convenient form:
\begin{equation} 
 A_{ep0-}=\frac{\eta_p x_e (1+x_p) - \eta_e x_p (1+x_e) }{\eta_p x_e (1+x_p) + \eta_e x_p (1+x_e)},  
\label{expasym}
\end{equation}
\noindent 
where $\eta =C_{max}/B_{max}$ is a ratio of the measured Compton and 
Bremsstrahlung edges, $x$ is the kinematic factor with indexes $e$ and $p$ 
denoting the measurement with electrons and positrons respectively.  

Uncertainty of the factor 
\hbox{$x_{e,p}=4{\cal E}_{e,p} \omega_0\sin^2{(\theta_0/2)}/m^2$}
is dominated by the lepton beam energy spread while contribution of the
other constituents is negligible: 
$\sigma (\omega_0)/\omega_0\approx 10^{-5}$,
$\sigma (m)/m \approx 3\cdot10^{-7}$,
$\Delta (\theta_0)\!\approx\!2~mrad\Rightarrow$ 
$\Delta\sin^2{(\theta_0/2)}\approx 3\cdot10^{-6}$. 
Therefore,  HERA leptons' energy spread
$\sigma ({\cal E})/{\cal E}\approx 10^{-3}$ will introduce a 
systematic uncertainty $7.9\cdot10^{-4}$ to the measured asymmetry.
Since the electron and positron measurements are considerably separated in time,
one of important potential sources for a false asymmetry could be
detector aging and degraded performance. This assumption could be checked by measuring
the calorimeter resolution from the spectra edges' slops - the same  energy
resolution $\sigma_0 = 0.24$ for both spectra indicates the same, non-degraded, 
response of the calorimeter.

A major energy correction factor, detector 
non-linearity, applied to each measured Compton edge will largely cancel 
in the asymmetry. 
However,  assuming a possible change of detector's
linearity we apply a non-linearity correction 
(as described in ref.~\cite{Gharibyan:2003fe})
which will scale the ratio
$\eta \to \eta (1+f {\cal E}(1-\eta))$ with a factor 
$f=0.001 GeV^{-1}$. 
Limiting the possible change of non-linearity factor
to 10\%, from error propagation in the Eq.(\ref{expasym}),
we derive an associated conservative  systematic 
uncertainty of $9.6\cdot10^{-4}$ for the asymmetry which is affected
negligibly by the correction.

%%----------------------------------------

A correlated error could potentially be induced by the detector spatial 
non-uniformity which affects the Compton edge absolute energy determination 
in case of Compton and Bremsstrahlung beams vertical separation~\cite{Gharibyan:2003fe}. 
In the asymmetry, however, the non-uniformity is negligible when the 
Bremsstrahlung from electrons and positrons overlap  (Eq.(\ref{expasym}) 
effectively depends on 
$$\eta_p / \eta_e = (C_{max}^p /C_{max}^e )\cdot (B_{max}^e /B_{max}^p)$$ ratio). 
This is the usual case with the calorimeter centered on the Compton beam during 
polarization measurements.
Then, the electron and positron Bremsstrahlung vertical overlap is provided by geometrical 
constraints, flat orbit requirements and the polarimeter setup location at HERA West –- 
away from the routine beam-tuning regions at the lepton-proton interaction points in 
HERA East, North and South.  
In order to depart from the assumed $100\%$ spatial correlation of $e^-$ and $e^+$ 
Bremsstrahlung beams, we allow a $10\%$ mismatch (correlation coefficient $C_f =0.9$) to estimate 
the detector non-uniformity influence on the measured asymmetry. Thus, the quoted $1\%$ 
spatial non-uniformity~\cite{Barber:1992fc} will result in a $2.2\cdot10^{-3}$ 
correlated systematic error in the asymmetry
according to correlated error propagation formula
\begin{equation}
(\delta A)^2 = (dA_p)^2+(dA_e)^2-2C_f dA_p dA_e,
\label{correrr}
\end{equation} 
\noindent 
with \hbox{$dA_p=\frac{\partial A}{\partial\eta_p}\Delta\eta_p$},
\hbox{$dA_e=\frac{\partial A}{\partial\eta_e}\Delta\eta_e$} and 
\hbox{$\Delta\eta_p/\eta_p=\Delta\eta_e/\eta_e=0.01$}. 

%%     -----------------------------
Other minor instrumental false asymmetry sources are electronic 
pedestal offsets, estimated to be $\pm 4MeV$ with $\approx 3\cdot10^{-4}$
contribution to asymmetry and luminosity dependent pile-up photons. 
The latter amounts to 0.02 and 0.0015 photons/bunch 
for the electrons and positrons respectively, and, in our case,
will tend to reduce any true asymmetry so, it can't be responsible 
for the observed effect. An ignorable, about the order of m, contribution 
to electron-positron asymmetry is coming from Bremsstrahlung edge charge 
dependence.    
%%%%%%%%%%%%%%------------+++++++++++++++++2222222222222222
The quoted and calculated errors are listed in the Table~1.

\begin{table}[h]
\centering
\parbox{\hsize} 
\caption{\label{tab1} { Table 1: Errors of the measured asymmetry $A=1.3\cdot 10^{-2}$. }} 

\vskip 3mm
\begin{tabular}{l | c | c}
\toprule
Source & Magnitude (rel.) & Error $\Delta A$   \\
%\hline
\toprule
Statistical fluctuations & $1.4\cdot 10^{-3}$ & $1.3\cdot 10^{-3}$ \\
Laser frequency shift& $10^{-5}$ &$\approx 0$\\
Electron mass' error & $3\cdot10^{-7}$ &$\approx 0$\\
Interaction angle drift & $3\cdot10^{-6}$ &$\approx 0$\\
HERA-e energy spread& $10^{-3}$ & $7.9\cdot 10^{-4}$ \\
Non-linearity change& $10^{-1}$ & $9.6\cdot 10^{-4}$ \\
Beams vertical mismatch& $10^{-1}$ & $2.2\cdot 10^{-3}$ \\
Electronic pedestal shift & $2.9\cdot 10^{-4}$ & $3\cdot 10^{-4}$ \\
Pile-up photons& $7.5\cdot 10^{-2}$ &$\le 0$\\
$e^+$ bremsstrahlung edge& $7.2\cdot 10^{-5}$ &$\approx 0$\\
%\hline
\toprule
\end{tabular}
\end{table}

%%----------------------------------------------
{

There are also potential contributions from chiral sources collected in 
the Table~2. To estimate the magnitudes of these effects I have explored 
the HERA transverse polarimeter setup simulation code~\cite{Gharibyan:2006note}.

\begin{table}[h]
\centering
\parbox{\hsize} 
\caption{\label{tab2} { Table 2: Systematic contributions from chiral sources. }} 
\vskip 3mm
{ \begin{tabular}{l c c}
\toprule
Source                           & Magnitude  & Error $\Delta A$   \\
\toprule
Magnetic field & $50~\mu T$ & $2.3\cdot 10^{-26}$ \\
Earth's spin & $2\cdot 10^{-23}~eV$ & $8.0\cdot 10^{-30}$ \\
Lepton polarization & 0.6 & $3.1\cdot 10^{-6}$ \\
Laser linear polarization & 0.2 & $5.4\cdot 10^{-7}$ \\
\toprule
\end{tabular}}
\end{table}

A tiny effect from the stray magnetic fields have been calculated using 
Eq.(\ref{emref}). The Earth's rotation gravitomagnetic impact on 
elementary particles amounts to
\hbox{$\hbar g/c=2.2\cdot 10^{-23}$eV},
where $g$ is the gravitational acceleration in the laboratory~\cite{Obukhov:2000ih}.
This contribution to the measured asymmetry is also negligibly small.
Compared to the birefringent magnetic field and Earth's spin, the instrumental chiral effects, 
responsible for false asymmetries, are larger by many orders of magnitude. 
The effects are induced by a change of laser polarization either linear or circular,
the latter in convolution with the lepton spin. An energy dependence of the induced
asymmetries (for 100\% polarized leptons) is plotted in Fig.4 for HERA polarimeter 
parameters. Though the mentioned polarized effects can not directly affect the 
Compton edge photons to alter the observed asymmetry, an energy smearing and 
spatial non-uniformity of the detector can mimic a Compton edge shift.
Simulations with conservatively high values of the lepton and linear light 
polarizations show artificial asymmetries below $10^{-5}$ value.

}
\color{black}{}

%%%****************************************************************************888888888888

\begin{figure}[ht!]
\centering
\includegraphics[scale=0.6]{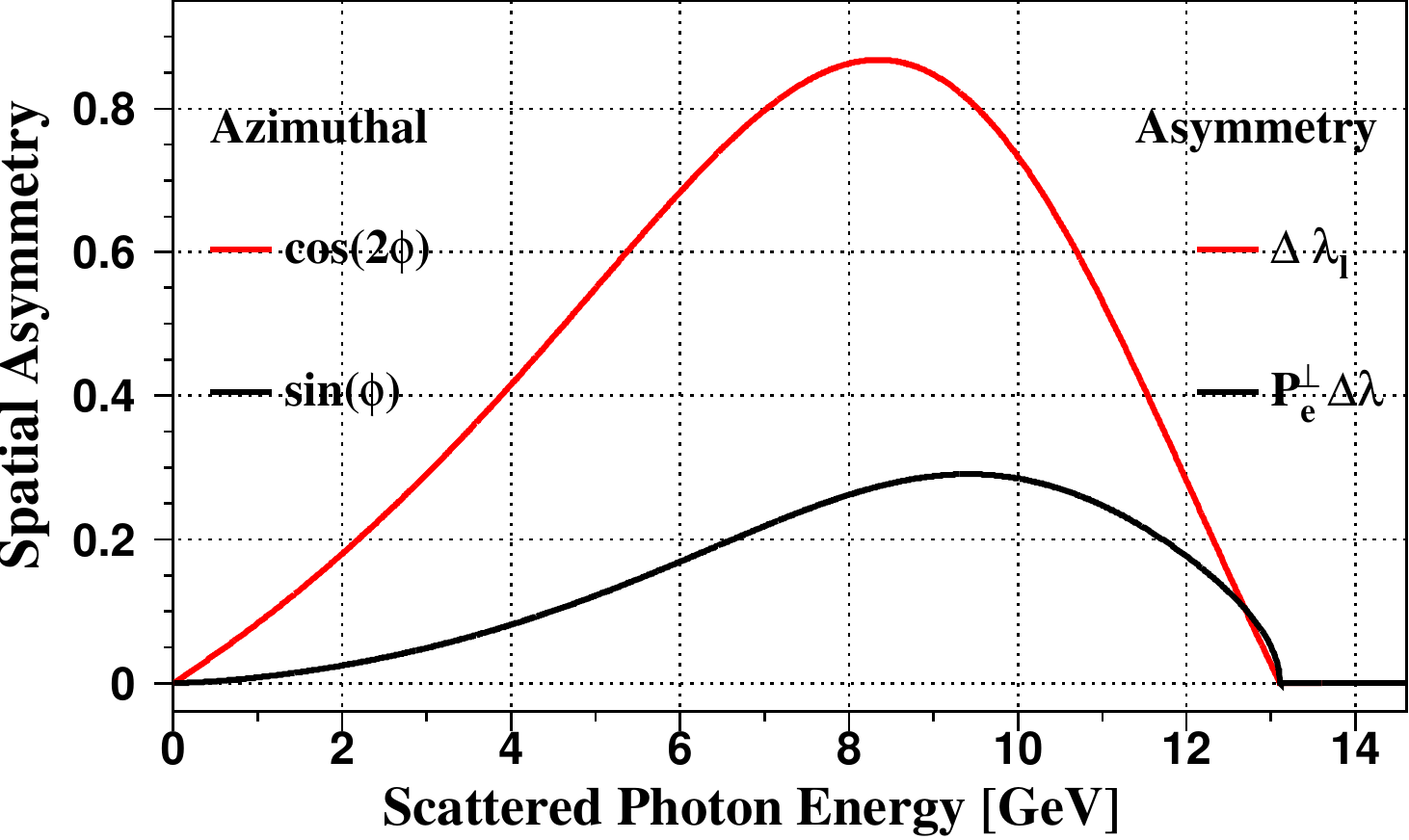}
\caption{\label{fig4}
{
Asymmetries in Compton scattering 
associated with the initial photon's polarization (linear $\lambda_l$ 
and circular $\lambda$) changes.
Analyzing powers (asymmetries for the cases 
\hbox{$
\lvert \Delta\lambda_l\rvert=1,
\lvert P_e \Delta\lambda\rvert=1
$}) of the 
HERA transverse polarimeter setup.}
}
\end{figure}

%%%%%%%%%%%%%%------------+++++++++++++++++2222222222222222
With a quadratic sum of the estimated systematic and statistical errors 
the measured asymmetry and its error become
\begin{equation}
 A_{ep0-}=(1.293 \pm 0.286)\cdot 10^{-2},
\label{resultasym}
\end{equation}
which differs from zero by more than $4\sigma$ confidence.
Within an extreme conservative approach, a linear summation of all 
the errors reduces the deviation to $3.6\sigma$ confidence interval.

Separate contributions
of the gravitational charge and space parity violations can't be derived from 
this asymmetry alone.
Instead one can assume space P parity conservation and calculate maximally
possible electron-positron charge C parity violation by 
inserting the observed asymmetry into  Eq.(\ref{dgog}):
\begin{equation}
\Delta U_{Gp}=(-9.0\pm 2.0)\cdot 10^{-12}.
\label{result0}
\end{equation}
In a similar way, within a perfect C parity conservation
one will have maximal P parity violation magnitude using 
Eq.(\ref{dgog_}):
\begin{equation}
\Delta U_{G-}=(1.28\pm 0.29)\cdot 10^{-11}.
\label{result1}
\end{equation}
In a general case of  C\&P violation 
the C and P symmetries are violated to lesser degrees.
Obtained signs of violations correspond to a stronger gravitational coupling
for the left helicity electrons  relative to the right helicity positrons.

\section{Conclusions}

Applying a gravitational field-induced refraction 
and assuming an equivalence principle violation in a general form 
$\Delta G_p /G$ for  positrons and $\Delta G_-/G$ for helicty, an outstanding
sensitivity has been demonstrated for the high energy Compton scattering 
to such gravitational anomaly. 
Within the developed formalism, the HERA Compton polarimeter's 
recorded spectra with electrons and positrons strongly disfavor the positron's 
anti-gravity and show a significant deviation of the $\Delta G_p /G$ or 
$\Delta G_-/G$ from zero.
The last claim is based on a detected $0.013$ energy asymmetry, which is a large 
number compared to the laser and lepton beam energy relative uncertainty of 
$10^{-5}$ and $10^{-3}$, respectively. The remaining  source of a
possible systematic energy error is the detector that is greatly eliminated 
from final result by using the asymmetry instead of absolute energy measurements.  
However, additional uncorrelated systematic errors  may impair the outcome 
and, claiming a definite observation of CP parity violation at high energy 
gravitational interactions would require the following:

-- a thorough analysis of many Compton spectra 
accumulated and recorded by the HERA during its running period; 

-- elimination of possible electroweak sources  that can mimic such result;

-- experimental verification at other accelerators.

\noindent In the absence of these, the measured electron-positron asymmetry 
can only be considered as a hint for the gravitational symmetry breaking and an 
invitation  for further studies. 
New experiments, however, will require future $e^-e^+$ machines with sufficiently 
high $\gamma$ or a precise setup on the currently running 6~GeV accelerator 
PETRA-III with the highest positron energy available.
Anyway, it is worth the efforts since high energy violation of the equivalence
principle and gravitational CP parity 
could reveal an interaction to massive or lower spin gravitons. 

\section*{Acknowledgement}
I thank B.~Sobloher and S.~Schmitt for providing details about the
positron generated spectra, and R.~Brinkmann for details about the electron 
measurement and the HERA. I'm thankful also to 
 A.~Buniatyan and K.~Balewski for useful discussions.

\end{document}